\documentstyle[12pt]{article}
\begin{document}
\begin{titlepage}
\hfill TUW--96--13 \\[1.0cm]

\begin{center}
{\LARGE
General Treatment of All 2d Covariant Models
\footnote{Invited Lecture at XII International Hutsulian Workshop 
of ``Methods of Mathematical Physics'', Rakhiv, Ukraine, Sept.\ 
11-17, 1995}}\\[2.5cm]

W.\ Kummer\footnote{wkummer@tph.tuwien.ac.at}\\
Institut f\"ur Theoretische Physik \\
Technische Universit\"at Wien\\
Wiedner Hauptstr. 8-10, A-1040 Wien\\
Austria\\
\end{center}
% \vfill
\vspace{2.5cm}
\begin{abstract}
General matterless models of gravity include dilaton gravity, 
arbitrary powers in curvature, but also dynamical torsion. They 
are a special class of "Poisson--sigma--models" whose solutions 
are known completely, together with their general global 
structure. Beside the ordinary black hole, arbitrary singularity 
structures can be studied. It is also possible to derive an 
action "backwards", starting from a given manifold. The role of 
conservation laws, Noether charge and the quantization have been 
investigated. Scalar and fermionic matter fields may be included 
as well.
\end{abstract}
\vfill
\end{titlepage}

\newpage

\section{Introduction} 
The interest in two dimensional diffeomorphism invariant theories has many 
roots. Presumably the most basic one is the central role played by 
spherically symmetric models in d = 4 General Relativity (GR) as a 
consequence of Birkhoff's theorem. Two dimensional models with 'time' and 
'radius' possess an impressive history with promising recent developments 
\cite{ber}. On the other hand, the fact that the Einstein--Hilbert action of pure 
gravity in 1+1 dimensions is trivial, also has spurred the development of 
models with additional nondynamical \cite{tei} and dynamical (dilaton and tachyon) 
scalar fields \cite{bro}, besides higher powers of the curvature 
\cite{bro,ban}.
Especially the study of 2d--dilaton theories turned out to be an important 
spin--off from string theory, and has led to novel insights into properties 
of black holes \cite{bro,ban,man}. For a scalar field, coupled more generally than a 
dilaton field, even more complicated singularity structures have been found 
\cite{lem} than in the ordinary dilaton--black hole \cite{bro}. \\
Actually such structures were known already before in another branch of 
completely integrable gravitational theories which modify 
Einstein--relativity in 1+1 dimensions by admitting nonvanishing dynamical 
torsion \cite{kata,katb}. Here the introduction of the light--cone (LC) gauge led to 
the expression of the full solution in terms of elementary functions 
\cite{kuma} and 
to an  understanding of quantum properties of such a theory in the 
topologically trivial \cite{kumb} and nontrivial \cite{sch} case. \\
Recently important progress has been made by the insight \cite{stra} that
\underbar{all} theories listed above are but special cases of a
'Poisson--sigma--model' (PSM) with action 
\begin{equation} 
    L = \int\limits_{M} (A_{B} \wedge d X^{B} + \frac{1}{2} P^{B C}(X) A_{B}
      \wedge A_{C})\;\; .  \label{Gl:1}
\end{equation}
The zero forms $X^B$ are target space 'coordinates' with connection one 
forms $A_B$. $P$  expresses the (in general degenerate) Poisson--structure on 
the manifold $M$, it has to obey a Jacobi--type identity, generalizing the 
Yang--Mills case, where $P$ is linear in $X$ and proportional to the structure 
constants. For the subclass of models describing 2d--covariant theories the 
$A_B$ are identified with the zweibein $e^a$, with the connection 
${\omega^a}_b  = {\epsilon^a}_b\,\omega$,  and may include possibly further
 Yang--Mills 
fields  $A_i$. Introducing the Minkowskian frame metric $\eta_{ab} = 
{\rm diag}\,(1,- 1)$,  target coordinates $X^A$ on the manifold 
will be denoted  as  $ \{X^a,X,X^i\}$ .\\
Then the (matterless) dilaton, torsion, $f(R)$--gravity theories  and even 
spherically symmetric gravity are obtained 
as special cases \cite{strb} of an action of type (1), namely ($ \epsilon = 
\frac{1}{2} \varepsilon_{ab}e^a\wedge e^b $ )
\begin{equation}
     L = \int\limits_{M} (X_{a} D e^{a} + X d \omega - \epsilon V)\;\; 
     .\label{Gl:2}
\end{equation}
Appropriate fixing of $V = V(X^a X_a, X,Y)$ yields all the models listed 
above (and many more). It is possible in principle to write down the full 
solution for (\ref{Gl:2}) in an arbitrary gauge (coordinate system). As seen
below, the solution has much of the shape of the LC gauge solution
\cite{kuma}. \\
A crucial role for the integrability of theories (\ref{Gl:1}) in the general case 
play 'Casimir--functions'  $C_i(X^A)$ \cite{stra,strb} which on--shell become 
constants and thus (gauge--inde\-pendent) 'observables',  also in the classical 
case \cite{int}. These constants $C_i$ together with other parameters in $P$ (or
$V$, e.g.\ the cosmological constant) determine the almost limitless variety of 
Penrose diagrams, characterizing the singularities of such theories 
\cite{klo}.  E.g.\ Schwarzschild or Reissner--Nordstr\"om black
holes  are just relatively simple members of that set. Also $C_1$ for the special 
theory \cite{kata} can be related to a global symmetry \cite{klo}. 
The relation to quasilocal energy \cite{yor} on a 'surface' and 
to Noether charge \cite{wal} has been clarified in 
\cite{kume}.\\  

Furthermore it is possible to show how a theory quadratic in torsion and curvature may indeed be
reformulated as an equivalent dilaton theory, 
by a {\sl local} method starting from a first order formalism for the
$R^2+T^2$--theory \cite{katc}.  Here the spin connection is
eliminated in favor of the torsion which turns into a nondynamical
field variable. \\

\section{PSM--Gravitation}
\subsection{The General Model}

With $ V $ in (\ref{Gl:2}) depending linearly on $ X^{a} X_{a} $
\begin{equation}
   V = \frac{\alpha}{2} X^{a} X_{a} + v(X,Y)\;\; ,
     \label{Gl:3}
\end{equation}
the equations of motion from (\ref{Gl:2}) in a LC basis of the frame metric 
( $ \varepsilon_{+-} = -1$ , ${ X^{\pm} = (X^0 \pm X^1)/\sqrt{2}}$,
$ \eta_{+-} = \eta_{-+} = 1 $) are
\begin{eqnarray}
     d X^{\pm} \pm \omega X^{\pm} &=& \pm e^{\pm} V \nonumber\\
     d X + X^{-} e^{+} - X^{+} e^{-} &=& 0 \label{Gl:4}
\end{eqnarray}
and
\begin{eqnarray}
  d e^{\pm} \pm \omega \wedge e^{\pm} = - \alpha e^{+} \wedge e^{-} X^{\pm} \nonumber \\
  d \omega = - e^{+} \wedge e^{-} \frac{\partial v}{\partial 
  X}\;\; .  
  \label{Gl:5}
\end{eqnarray}
Multiplying the first pair of equations in (\ref{Gl:4}) with $X^-$ and $X^+$, 
respectively, the second one with V and adding yields 
\begin{equation}
   d(X^{+}X^{-}) + V d X = 0, \label{Gl:6}
\end{equation}
producing an absolutely conserved quantity ($d\, C = 0$)
\begin{eqnarray}
    C_{1}= C =  X^{+}X^{-} e^{\alpha X} + w(X) 
   \label{Gl:7}\\
   w(X) = \int\limits_{X_{0}}^{X} v(y) e^{\alpha y} dy\;\; . 
   \label{Gl:8}
\end{eqnarray}
Clearly the lower limit $X_0 = const.$ must be determined appropriately so that 
(inside a certain patch) the integral exists for a certain range of  
 $X$. Eq.\ (\ref{Gl:7}) generalizes \cite{stra,strb} the previously known 
\cite{kata,kuma} analogous quantity for 2d gravity with dynamical
torsion. However,  the limit $\alpha \to 0$ immediately also yields the
conservation law for torsionless cases (F(R)--gravity, dilaton gravity
\cite{bro} etc.). \\
Setting one LC component of the torsion, e.g.\ $X^+$ identically zero, the 
first equation (\ref{Gl:5}) (at $e^+ \neq 0$) may yield a constant 
curvature.   Such 'de--Sitter' 
solutions are related to $ C = 0$ (within an appropriate convention for the 
integration constant in (\ref{Gl:8})). Representing discrete points in phase 
space they are notorious  especially in the quantum case 
\cite{sch}. If $X^+ \neq 0$ 
the solution of (\ref{Gl:4}) and (\ref{Gl:5}) becomes 
\begin{eqnarray} \label{Gl:9}
     e^{+}  & = & X^{+} e^{\alpha X} d f\nonumber \\
     e^{-}  & = & \frac{d X}{X^{+}} + X^{-} e^{\alpha X} d f \\
     \omega & = & - \frac{d X^{+}}{X^{+}} + V e^{\alpha X} d 
     f\;\; . \nonumber
\end{eqnarray}
Of course, for $X^- \neq 0$ the analogous solution exists with the
roles of $ X^+ \leftrightarrow X^-$ exchanged.  The first terms in
the first three eqs.  for $ e^{\pm} \rightarrow \delta e^{\pm},
\omega \rightarrow \delta \omega, d f \rightarrow \delta \gamma $ are
the on--shell extension of a global nonlinear off--shell symmetry of
(\ref{Gl:2}), \cite{kumc}.  It is related to the conservation
$\partial_\mu\, J^\mu_\nu = 0$ of a Noether current $ J^\mu_\nu =
C\,\delta^\mu_\nu$ because under such a transformation the Lagrangian
density in (2) changes by a total derivative only.  Mathematically
(\ref{Gl:9}) coincides with the solution in the LC--gauge \cite{kuma}
where the curvature $X$ is gauge--fixed to be linear in 'time'.  But
(\ref{Gl:9}) has the big advantage that it is valid in an arbitrary
gauge, whereas solutions obtained in the literature to such theories
had to rely on special gauges and sometimes on sophisticated
mathematical methods to solve the respective equations (cf.\ e.g.\
\cite{ban,man,lem,kata}).  The line element from
(\ref{Gl:9}) generally reads

\begin{equation}
   (ds)^{2} = 2 e^{\alpha X} d f \otimes (d X + X^{+} X^{-} e^{\alpha X} d 
   f)\;\; ,\label{Gl:10}
\end{equation}    
with $X^+X^-$ to be expressed by (\ref{Gl:7}) for fixed $C$. For the case with 
torsion a generic model  may be chosen as
\begin{equation}
   V = \alpha X^{+}X^{-} + \frac{\rho}{2} X^{2} - \Lambda\;\; . 
   \label{Gl:11}
\end{equation}
This $V$ allows to produce $C$ by a simple integral according to 
(\ref{Gl:7}).
Integrating out $X$ and $X^\pm$ in (\ref{Gl:2}) leads to
the model quadratic in curvature and torsion of \cite{kata,kuma} which, in
four dimensions, together with the Einstein--Hilbert term has been known as
the 'Poincare--gauge theory' for some time \cite{he}. It only contains second 
derivatives in the field equations for the variables $e^a$ and $\omega$. 
However, higher derivative theories are to be treated with equal ease, when 
polynomials of higher degree in $X$ and $X^+X^-$ are admitted in 
(\ref{Gl:3}). Of 
course, $V$ could even be a nonpolynomial function. This 
would only make the integration harder which leads to (\ref{Gl:7}). As we shall recall 
shortly below, the zeros of (\ref{Gl:7}) determine the singularity structure of
the theory. Thus one could design such a structure by prescribing  
$ C (X^{A})$ . 
>From (\ref{Gl:7}) the corresponding $V$ can be read off by differentiation, and the 
action for that structure follows immediately. \\
Among the models with vanishing torsion ($\alpha = 0$ in (\ref{Gl:3})), the 
Jackiw--Teitelboim model obtains for $v = \Lambda X$. Witten's black hole 
\cite{bro} represents a special case of a class of more general torsionless 
theories involving the curvature scalar R and one additional scalar field 
\cite{ban} in a Lagrangian of the type 
\begin{equation}
    {\cal L} = \sqrt{-g}[ \partial_{\alpha} \varphi \partial_{\beta} \varphi
               g^ {\alpha\beta} + A(\varphi) + R B(\varphi)] 
               \label{Gl:12}
\end{equation}
with arbitrary functions A and B. Matterless dilaton--gravity \cite{bro} is
the special case $\varphi^2 = 4B = A/\lambda^2 = 4\,e^{-2\Phi}$
\begin{equation}
    {\cal L}_{dil} = \sqrt{-g} e^{- 2 \Phi} [4 \partial_{\alpha} \Phi
      \partial_{\beta}  \Phi g^{\alpha \beta} + 4 \lambda ^{2} + R]\;\; .
      \label{Gl:13}
\end{equation}
Using the conformal identity for $\tilde{g}_{\alpha\beta} = e^{-2\phi} 
g_{\alpha\beta}$ (or $\tilde {e}^a = e^{-\phi}\, e^a$)
\begin{equation} 
     \sqrt{-\tilde{g}} \tilde{R} = \sqrt{-g} R + 2 \partial_{\alpha}
     (\sqrt{-g} g^{\alpha \beta} \partial_{\beta}\phi)\;\; . 
     \label{Gl:14}
\end{equation}
Eq.\ (\ref{Gl:14}) allows the elimination of the kinetic term for $\varphi$ in 
(\ref{Gl:12}). The resulting action may be written readily  in the
first order  form (\ref{Gl:2}) for $\tilde e^a = e^{-\Phi} e^a, V = +4 \lambda^2, 
X = 2 e^{-2\Phi}$. Going back from (\ref{Gl:9}) for $\alpha = 0$ as 
$ e^a = \tilde e^a / \sqrt{X/2}$, simply leads to
\begin{eqnarray}
      e^{+} & = & X^{+} e^{\Phi} d f\nonumber \\
      e^{-} & = & \frac{1}{X^{+}}[-4 d \Phi e^{-\Phi} + d f (C\, e^{ \Phi}
                   - 8 \lambda^{2} e^{-\Phi})] \label{Gl:15}\\
      \omega & = & - \frac{d X^{+}}{X^{+}} + 4 \lambda^{2} d f\;\; . \nonumber
\end{eqnarray}
Here $\Phi, f$ and $X^+$ are arbitrary functions. E.g.\ the Kruskal form for 
the metric $(ds)^2 = 2e^+ \otimes e^-$ a dilaton black hole follows from the 
gauge--fixation $(X^+X^- = uv)$
\begin{eqnarray*}
     8 \lambda^2 e^{-2 \Phi} & = &  C - uv \\
     4 \lambda^{2} f & = &  \ln\, u\;\; .
\end{eqnarray*}
The mass of the dilaton black hole is related to $C$ by $ C = 8 \lambda M $.
$X^+ \neq 0$ being still arbitrary, it may be used to gauge $\omega = 0$ 
which shows that the connection $\omega$ in (\ref{Gl:15}) really has nothing to
do with a curvature belonging to the metric derived from that equation. \\
Now precisely the same procedure may be applied to the generalized theories of 
type (\ref{Gl:12}). Here $\varphi$ can be eliminated \cite{ban} using
(\ref{Gl:14}) 
\begin{equation}
  \tilde{{\cal L}}= \sqrt{-{\tilde g}}[A(\varphi)/F(\varphi) + \tilde{R} B(\varphi)]
                                          \label{Gl:16}
\end{equation}
with 
\begin{eqnarray}
     g_{\alpha \beta} & = & \tilde g_{\alpha \beta} / F(\varphi) \nonumber\\
     \ln\, F(\varphi) & = &  \int\limits^{\varphi} dy/(\frac{d B}{d 
     y}) \;\; .\label{Gl:17}
\end{eqnarray}
The corresponding first order action (\ref{Gl:2}) becomes
\begin{eqnarray*} 
   \tilde{L} = \int (X_{a} D \tilde{e}^a + 2 B d \omega - \tilde{\epsilon} 
   A/F)\;\; .   \nonumber
\end{eqnarray*}
In (\ref{Gl:2}) for $\alpha = 0$ we have as a consequence

\parbox{12.0cm}{
\begin{eqnarray*}
    X & = & 2 B \\
    V = v & = &  \frac{A(B^{-1}(X/2))}{F(B^{-1}(X/2))} \;\; ,
\end{eqnarray*}}\hfill\parbox{1.0cm}{\begin{eqnarray}\label{Gl:18}
\end{eqnarray}}\\
and the conserved quantity for any theory of type (\ref{Gl:12}) is 
(\ref{Gl:7}) with 
$\alpha = 0$ and
\begin{equation}
    w(X) =  \int\limits_{B^{-1}(X_{0}/2)}^{B^{-1}(X/2)} \frac{A(y) d 
    y}{F\,(y)}\;\; .\label{Gl:19}
\end{equation}
The line--element in terms of coordinates $(f,X) = Y^\alpha$  for any
such theory and in any gauge reads
\begin{eqnarray}
        (d s)^{2} = F^{-1} 2 d f \otimes [dX + df ( C - w(X))]\;\; 
        .\label{Gl:20}
\end{eqnarray}

Of course, in each application to a particular model a careful analysis of 
the range of validity of the mathematical manipulations is required in order 
to determine a patch, where those steps are justified: allowed ranges for 
transformations of fields, inversions of functions like $B^{-1}$, 
integrability of $F$, admissible gauges for $f$ and $X$ etc. \\
Another example is the action for the Schwarzschild black hole in 4d GR. 
$v(X) = - 1/(2 X^2)$ in (\ref{Gl:2}) is found to yield the correct 
line--element .\\

\subsection{Killing Vector and Singularities}

In our very general class of models the Killing vector can be found  
without fixing the gauge (coordinate--system). 
Using (\ref{Gl:9}) we rewrite the line element  (\ref{Gl:10}) in a 
theory (\ref{Gl:2}) as 
\begin{equation}
     (d s)^{2} = d f \otimes [ 2 e^{\alpha X}d X + l d 
     f]\label{Gl:21}
\end{equation}
where
\begin{equation} 
    l = 2 X^{+} X^{-} e^{2 \alpha X} = 2 e^{\alpha X}(C - w(X,Y))\;\; .
    \label{Gl:22}
\end{equation}
In terms of the variables $Y^\alpha = (f,X)$, resp. $\partial / \partial 
Y^\alpha , k^\alpha $  is the Killing vector with norm (\ref{Gl:22}) 

\begin{eqnarray} 
   k^\alpha &=& (1,0)\nonumber\\
   k^2 &=& k^{\alpha} k^{\beta} g_{\alpha \beta} = l\;\; .
   \label{Gl:23}
\end{eqnarray}
For the discussion of the
singularity  structure of (\ref{Gl:9}) a (partial) gauge fixing is useful. If
$ l > 0 $ in (\ref{Gl:21}) we choose coordinates time $(t)$ and space $(r)$ in $ f =
f(t,r), X = X(r)$ with $\dot f = T(t)$ and
\begin{equation}
  X' e^{\alpha X} + f' l(X) = 0\;\; ,\label{Gl:27}
\end{equation}
where $f' = \partial f / \partial r \;\; ,$ \quad $ \dot f = \partial f / \partial
t$. Introducing 
\begin{equation}
   K(z) = - \int\limits_{z_o}^{z} d y e^{\alpha y} l ^{-1} (y)\;\; ,
   \label{Gl:28}
\end{equation}
(\ref{Gl:27}) implies
\begin{equation} 
   f = \int\limits^{t} T(t') d t'+ K(X(r))\;\; .\label{Gl:29}
\end{equation}
In such a gauge $g_{t r}$, the off--diagonal part of the metric 
vanishes, so that `space` and `time` are  separated.  
In order to  avoid zeros in the norm of the Killing--vector field $k$ it is
obvious to  restrict $z$ and $z_0$ to a suitable interval of $y = X(r)$
where $k$  exists. The remaining elements of $g_{\alpha \beta}$  are:
\begin{eqnarray}
  g_{t t} &=& \dot{f}^{2} l\nonumber \\
  g_{r r} &=& - (f')^2 l\label{Gl:30}
\end{eqnarray}
Requiring a 'Schwarzschild'--form of the metric, i.e. $\det g = -1$, 
eliminates the arbitrary functions $T(t)$ and $X(r)$ altogether,
\begin{eqnarray}
   T &=& 1\nonumber \\
  \alpha\, X &=&  \ln\,(\alpha r)  \;    (\alpha \ne 0)\label{Gl:31}\\
   X &=& r \quad (\alpha = 0)\;\; ,\nonumber
\end{eqnarray}
dropping a multiplicative constant $a$ in $f$, and $1/a$ together with $r$,
and two further constants for the zero points of $t$ and $r$. Now 
\begin{equation}
  g_{t t} = - g_{r r}^{-1} = l(X(r))\label{Gl:32}
\end{equation}
follows with $ X(r) $ from (\ref{Gl:31}). Especially (\ref{Gl:32}) clarifies the
remark  above, how an action may be reconstructed for a given singularity in
the  metric, proceeding backwards through (\ref{Gl:22}) to (\ref{Gl:2}). \\
We note that for a (generalized) dilaton theory, besides $\alpha = 0$, 
because of the additional factor $1/F$ in (\ref{Gl:20}) there is a 
corresponding change to $l$ in (\ref{Gl:30}) etc. Thus the singularity 
structure is determined by $l / F$. \\
The Katanaev--Volovich model (\ref{Gl:9})  is 
sufficiently general to show the intrinsic singularity structure by an 
analysis of completeness of geodesics. $C^2$ global completeness was first 
shown in \cite{katb} within the conformal gauge. The more suitable 
present approach 
allows the extension to $C^\infty$ completeness and a discussion of possible 
compactifications \cite{klo}. In that model altogether 11 types of Penrose 
diagrams appear (G1, $\ldots$  G11 in the classification of \cite{katb}). 
Some show similarities to Schwarzschild and to Reissner--Nordstr\"om types, 
but there are many more.  In the more complicated cases they are obtained by
the  possibility to successively gluing together solutions for 
patches, each given by (\ref{Gl:15}). 
 The diffeomorphism for doing that is essentially
(\ref{Gl:28})  again. For further details we refer to \cite{klo}. 
 It is sufficient for our present purposes to note that for 
all types of singularities (including also e.g.\ naked ones) there are 
space--like directions allowing the study of surfaces (points) between such 
singularities at finite (incomplete case) or infinite (complete) distances. Also a
second point is obvious from this section: In all two dimensional theories the
conserved  quantity {\cal C} never has a well-defined sign. Thus any hope
to find a positive `energy` must be in vain. Therefore, also adding matter to
the theories (\ref{Gl:2}) is not likely to improve this situation.  

\section{Equivalence of Generalized Dilaton Theory}

We now show that in a theory of type (2), i.e. in a 'Poincar\'e--gauge theory' 
\cite{he} the torsion can be eliminated. In the definition
\begin{equation}
 \label{Gl:76}
 T^\pm = (\partial_\mu \pm \omega_\mu) e^\pm_\nu\, 
 \tilde\epsilon^{\mu\nu}
\end{equation}
we introduce light cone coordinates $T^\pm = \frac{1}{\sqrt{2}}(T^0 \pm 
T^1)$ in the Lorentz--indices. 
It should be noticed that the subsequent steps will even remain correct for 
$\alpha = \alpha (X)$ and 
general $v = v(X)$, i.e. a theory quadratic in torsion but with 
arbitrary higher powers in curvature. 

Now instead of $\omega_\mu$ the $T^\pm$ in (\ref{Gl:76}) are introduced as new 
variables: 
\begin{equation}
\label{Gl:77}
\tilde \epsilon^{\mu\nu} \partial_\mu \omega_\nu = \tilde 
\epsilon^{\mu\nu} \partial_\mu \tilde\omega_\nu + \tilde 
\epsilon^{\mu\nu}\partial_\mu \left[\frac{(e_\nu^-T^+ + e_\nu^+ 
T^-)}{e}\right]
\end{equation}
The first term on the r.h.s. of (\ref{Gl:77}) is proportional to a torsionless  
curvature $\tilde R$, 
\begin{equation}
\label{Gl:78}
\tilde\epsilon^{\mu\nu} \partial_\mu \tilde\omega_\nu = - \frac{\tilde R 
e}{2}\quad .
\end{equation}
Inserting (\ref{Gl:77}) into (\ref{Gl:2}), after shifting the derivatives in the 
second term of (\ref{Gl:77}) onto X exhibits the nondynamical nature of $T^\pm$ which may 
be 'integrated out' by solving their (algebraic) equations of motion. With a 
definition of the dilaton field
\begin{equation}
\label{Gl:79}
\frac{X}{2} = e^{-2\phi}
\end{equation}
and after reexpressing the factors $e_\nu^\pm / e$ from the square bracket 
of (\ref{Gl:77}) in terms of the inverse zweibeins  combining them into 
$g^{\alpha\beta} = e^{+\alpha}e^{-\beta} + e^{+\beta}e^{-\alpha}$, the 
Lagrangian ${\cal L}^{(1)}$ in (2) is found to be equivalent to
\begin{equation}
\label{Gl:80}
{\cal L}^{(2)} = \sqrt{-g} \left[- e^{-2\phi} \tilde R + 8 \alpha\cdot 
e^{-4\phi}g^{\mu\nu} (\partial_\mu\phi)(\partial_\nu\phi) - 
v(2e^{-2\phi}) \right]
\end{equation}
In addition, using the identity
\begin{eqnarray}
\label{Gl:81}
g_{\mu\nu} &=& e^{2\varphi}\; \hat g_{\mu\nu}\nonumber\\
\sqrt{-g} R & = & \sqrt{-\hat g}\hat R - 2\partial_\alpha \left[ 
\sqrt{-\hat g} \hat g^{\alpha\beta} \partial_\beta \varphi\right]
\end{eqnarray}
which also represents a local transformation, allows to write down the most 
general dilaton theory equivalent to the $R^2 + T^2$--theory: 
\begin{equation}
 \label{Gl:82}
{\cal L}^{(3)} = \sqrt{-\hat g}\left[ -e^{-2\phi}\hat R + 4 \hat 
g^{\alpha\beta} 
(\partial_\alpha\phi)\left(e^{-2\phi}\partial_\beta\varphi + 
2\alpha e^{-4\phi} \partial_\beta\phi\right)
 - e^{2\varphi} 
v(2e^{-2\phi})\right]
\end{equation}
For $\varphi = - \phi$
\begin{equation}
\label{Gl:83}
{\cal L}^{(4)} = -\sqrt{-\hat g} e^{-2\phi} \left[ \hat R + 
4(1-2\alpha e^{-2\phi}) (\nabla \phi)^2 + v(2e^{-2\phi})\right]
\end{equation}
the deviation from ordinary dilaton theory ($\alpha = 0, v = 4\lambda^2$) is 
most obvious. Of course, the dilaton field may be eliminated altogether as 
well, if in (\ref{Gl:82}) (for constant $\alpha$)
\begin{equation}
\label{Gl:84}
\varphi = \varphi (\phi) = \alpha e^{-2\phi} = \frac{\alpha X}{2}
\end{equation}
is chosen. In that case it seems more useful to retain the variable X 
instead of $\phi$:
\begin{equation}
\label{Gl:85}
{\cal L}^{(5)} = - \sqrt{-\hat g}\; \left[\frac{X \hat R}{2} + e^{\alpha X} v 
(X)\right]
\end{equation}
Comparing (\ref{Gl:85}) to a torsionless theory  with $\alpha = 0$ but 
modified $v$, the difference now just resides in the additional exponential 
$e^{\alpha X}$.

\section{Global Solutions for Dilatonic Versions of Theories with Torsion}

The study of global properties for 2d theories is based upon the
extension of the solution which is known at first only in local
patches,  continued maximally to global ones.  The analysis uses
null--directions which become the coordinates of Penrose diagrams
which are sewed together appropriately.  The continuation across
horizons and the determination of singularities can be  based upon
extremals or geodesics.  The physical interpretation of an
extremal is the interaction of the space--time manifold with a
point like test particle, 'feeling' the metric $g_{\alpha\beta}$
through the Christoffel symbol \cite{kuma}.  After torsion has been 
eliminated, there is no ambiguity for our analysis which only has 
extremals at its disposal. That interaction with extremals, 
however, crucially depends on the choice of the 'physical' metric
to be used: the one computed from the $e_\mu^a$ of (1), or any
$\hat g_{\alpha\beta}$ which is a result of different field
transformations involving the dilaton field? Clearly the
torsionless dilaton theory (\ref{Gl:80}) has the same metric as 
(\ref{Gl:1}), e.g.\ 
the global analysis of \cite{katb,klo} applies directly and the
different types of solutions are exhausted by those studied there.
However, from the point of view of a 'true' dilaton theory, one
could argue as in Section 2 that with a redefined matrix as in 
(\ref{Gl:83}),
 $\hat g_{\alpha\beta} =
e^{2\phi}g_{\alpha\beta} = 2g_{\alpha\beta} / X$ has some physical
justification as well.  In fact, for Witten's black hole
$g_{\alpha\beta}$ is flat and the interesting (black hole)
singularity structure  just results from the factor $2/X$.  Now, in
the original $R^2 + T^2$--theory [7]  there are solutions
(G3) resembling e.g.\  the black hole but not completely: Their
singularity resides at light--like distances and they are not
asymptotically flat in the Schwarzschild sense.  Thus the factor
$2/X$ may well yield improvements on that situation. \\
Here we indicate the analysis of a generalized dilaton 
gravity\cite{katc} 
\begin{equation}
\label{Gl:86}
{\cal L} = \sqrt{-\hat g} e^{-2\phi} \left[\hat R + 
4(1-2\alpha e^{-2\phi}) (\nabla \phi)^2 + 2\beta e^{-4\phi} + 4\lambda^2 \right]
\end{equation}
which is obtained from  (\ref{Gl:2}) by taking
\begin{equation} \label{Gl:87}
X = 2e^{-2\phi}~~~~
g_{\mu \nu} = {\hat g}_{\mu \nu} e^{-2\phi}~~~~
\Lambda = -4\lambda^2
\end{equation}
and omitting an overall minus sign.
We need to consider only the cases for $\beta = \rho$ positive, negative 
or $0$. The absolute value of a nonvanishing $\beta$ may always be 
absorbed by rescaling $X$ and $\omega$ to $X \rightarrow \sqrt{\beta}X$ 
and $\omega \rightarrow \frac{\omega}{\sqrt{\vert\beta\vert}}$.  
Let us start with a positive value for $\beta$ e.g.\ $+2$. 
All global solutions are most easily obtained by the known general 
solution (9).
Using  (\ref{Gl:87}) and defining coordinates by $v = -4f$, $u = 
\phi$  in (\ref{Gl:10}) yields
the line element of the generalized dilaton Lagrangian 
(\ref{Gl:86})
\begin{equation}
\label{Gl:88}
(ds)^2 = g(u) \left( 2dvdu+ l(u) dv^2 \right)
\end{equation}
with
\begin{equation}
\label{Gl:89}
l(u)=\frac{e^{2u}}{8} \left(
C- g(u) \left( \frac{4e^{-4u}}{\alpha} -
\frac{4e^{-2u}}{\alpha^2}+ C_0 \right) \right) 
\end{equation}
\begin{equation}
g(u)= e^{2\alpha e^{-2u}},\label{Gl:90}
\end{equation}

\begin{equation}
\label{Gl:91}
C_0 =  \frac{2}{\alpha^3}
+ \frac{4\lambda^2}{\alpha} 
\end{equation} 
which automatically implies the convention for the  constant of 
integration in (\ref{Gl:7})
to be used in the following.  
The conformal gauge $(ds)^2 = F(\tilde {u}')\, d\tilde {u}'\, 
d\tilde {v}'$ in (\ref{Gl:88}) is obtained by 
'straightening' the null extremals
\begin{equation}
\label{Gl:93} v=const. 
\end{equation}
\begin{equation}
\label{Gl:94} \frac{dv}{du}=- \frac2l \mbox{ for all $u$ with $l(u) \neq 0$} 
\end{equation}
\begin{equation}
\label{Gl:95} u=u_0 = const \mbox{ for $l(u_0)=0$}
\end{equation}
by means of a diffeomorphism 
\begin{equation}
\label{Gl:96}
\tilde {u}' = v+f(u),~~~~\tilde {v}'= v 
\end{equation}
\begin{equation}\label{Gl:97}
f(u) \equiv \int^u \frac{2 dy}{l(y)}.
\end{equation}  
A subsequent one  $\tilde {v}' \rightarrow \tan \tilde 
{v}'$ 
and another  appropriately chosen one
for $\tilde {u}'$ produce  the Penrose diagram. It is valid  for a 
certain patch where (\ref{Gl:97}) is well defined. 
Clearly the shape of those diagrams 
depends crucially on the (number and kind of) zeros and on the asymptotic
behavior of $l(u)$. The analysis of all possible cases as described 
by the ranges of parameters $\alpha, C$ and $\lambda^2$ is 
straightforward, but tedious. Apart from $C_0$, defined in 
(\ref{Gl:91}), also

\begin{equation}\label{Gl:98}
C_1 = \frac{2}{\alpha^2}\; e^{2\alpha\sqrt{-\lambda^2}}\; 
(\frac{1}{\alpha} - 2\, \sqrt{-\lambda^2})
\end{equation}
plays a role for $C < C_0$ and $\lambda^2 < 0$, discriminating the 
possible cases with two zeros, with one double--zero and without 
zero in $l$, i.e. the presence of two nondegenerate or one 
degenerate killing--horizon. The qualitatively distinct cases for 
$\alpha > 0$ and $\alpha < 0$ are listed in (\ref{Gl:99}) and 
(\ref{Gl:100}):

\underline{{$\boldmath{\alpha > 0:}$}}
\begin{equation} \label{Gl:99}
\begin{array}{l}
D1^+:~~~C> C_0   \\
D2^+:~~~C= C_0,
~~~\lambda^2 < 0 \\
D3^+:~~~C= C_0,
~~~\lambda^2 \geq 0 \\
D4^+:~~~C< C_0,
~~~C> C_1,
~~~\lambda^2 <0 \\
D5^+:~~~C< C_0,
~~~C= C_1,
~~~\lambda^2 <0  \\
D6^+:~~~C< C_0,
~~~C< C_1,  \\
\end{array}
\end{equation}

\underline{{$\boldmath{\alpha < 0:}$}}
\begin{equation} \label{Gl:100}
\begin{array}{l}
D1^-:~~~C>C_0 ,~~~C \geq 0 \\
D2^-:~~~C>C_0 ,~~~C < 0 \\
D3^-:~~~C=C_0 ,~~~C \geq 0 \\
D4^-:~~~C=C_0 ,~~~C <0 ,~~~ \lambda^2 \geq 0 \\
D5^-:~~~C=C_0 ,~~~C <0 ,~~~ \lambda^2 < 0 \\
D6^-:~~~C<C_0 ,~~~C< C_1 \\
D7^-:~~~C<C_0 ,~~~C= C_1 \\
D8^-:~~~C<C_0 ,~~~C> C_1, ~~~C<0 \\
D9^-:~~~C<C_0 ,~~~C> C_1, ~~~C \geq 0\\
\end{array}
\end{equation}

For the corresponding Penrose diagrams we refer to \cite{katc}.  
Except for the cases $D2^+$, $D3^{\pm}$, 
$D4^-$, $D5^-$, $D6^-$ (where $R \to \pm\alpha\, C_0$)   the scalar 
curvature diverges at 
$u \rightarrow \pm \infty$. 
For each set of the parameters as summarized in (\ref{Gl:99}), 
 and (\ref{Gl:100}) 
another solution  in conformal coordinates is obtained by 
interchanging the role of the null--directions.
 The transformation

\parbox{10cm}{
\begin{eqnarray*}
\label{Gl:101}
\tilde{u}'' &=& u \\
\tilde{v}'' &=& -f(u) - w
\end{eqnarray*}}\hfill
\parbox{1cm}{\begin{eqnarray}\end{eqnarray}}\\
with $f(u)$ from (\ref{Gl:97}) may be easily verified to do this job. 
\\
Extremals obey simple first order differential equations. All cases 
(time--like, space--like, null) must be checked, especially at the 
boundaries. With these tools patches may be glued together.  For 
$D2^+$ this leads to the well--known shape of the 'classical' black 
hole in the corresponding global solution. It seems instructive 
to compare our present global structure to the 
one studied for other theories. The original $R^2+T^2$--theory 
contains one solution resembling the 'real' Schwarzschild black hole 
only in a very approximate sense. In the notation of \cite{katb}  
the solution $G3$ exhibits an (incomplete) singularity, but into 
null--directions, the 'asymptotically flat' direction is replaced 
there by a singularity of the curvature, albeit at an infinite 
distance (complete case). Here precisely the example $D2^+$ 
is completely Schwarzschild--like. 
 Other solutions with similar properties, but 
more complicated singularity structure are $D3^-, \; D4^-$ 
(naked singularities) and $D5^-$. Among the remaining 
diagrams the absence of manifolds with two dimensional (infinite) 
periodicity as in the $R^2+T^2$--case can be emphasized. On the 
other hand, the 'eye' diagram $D6^+$ appears here, as well as the 
square diagrams $D1^+$ of $R^2$--gravity \cite{lem}.  $D5^-$ 
represents an interesting variety of a manifold where the ordinary 
black hole is replaced by a 'light'--like singularity. 

\section{Conclusion and Outlook}

In the quantum case a genuine {\sl field} theory only arises in 
interaction with matter. Without that only on a suitable compactified space 
isotopic to $S^1$ the finite number of zero modes precisely of the $C$-s covers 
a quantum mechanical theory with a finite number of degrees of freedom. \\
Although $C_1$ (in our generic case) turns into a 'energy density', not 
necessarily constant in space and time anymore when matter is present 
\cite{kumd}, it retains its physical aspects related to the geometrical part of the 
action --- very much like the mass parameter in the so far very most 
prominent case, the example of the dilaton black hole interacting with 
matter: E.g.\ generalizing to a matter dependent (minimally coupled) 
action  $L^{(m)}(e_\mu^a)$ with a contribution to the r.h.s.\ of the first 
eq.\ (\ref{Gl:4}), containing a one--form ${\cal S}^{(m)\pm}$, the steps 
leading to (\ref{Gl:7}) imply a relation 

\begin{equation}
dC + W^{(m)} = 0 
\label{Gl:102}
\end{equation}
Poincar\'e's lemma or the other e.o.m-s (\ref{Gl:5}) require that
\begin{equation}
W^{(m)} = X^+ {\cal S}^{(m)-} + X^-{\cal S}^{(m)+} = d\, C^{(m)}
\end{equation}
Thus a generalization of the absolute conservation law for $C$ 
emerges.  $C$ now will vary in space and time, in general. 

All our results, however, point into the direction that in 1 + 1 
dimensions, including spherically symmetric gravity and thus also the 
Schwarzschild black hole, the geometrical part of the action does {\it 
not} acquire quantum corrections \cite{kumb}, such quantum effects 
being restricted to compactified topologies \cite{sch}.

\section*{Acknowledgement:}

The material presented here is mainly based upon collaborations 
with M.O.\ Katanaev, H.\ Liebl and P.\ Widerin. 
The author has also profited from discussions  with H.\ Balasin, 
T.\ Kl\"osch, P.\  
Schaller and T.\ Strobl. The results presented in this lecture have 
been supported by Fonds zur F\"orderung 
der wissenschaftlichen Forschung (Project P-10221--PHY). 
%% \newpage
%% \section {References}


\begin{thebibliography}{99}
\bibitem[1]{ber} D.K.\ Berger, D.M.\ Chitre, V.E.\ Moncrief and Y.\ Nuther, Phys.\ Rev.\ {\bf 
                D5} (1973) 2467 \\
                P.\ Thomi, D.\ Isaak and P.\ Hajicek, Phys.\ Rev.\ {\bf D30} (1964) 
                1168\\
                T.\ Thieman and H.A.\ Kastrup, Nucl.\ Phys.\ {\bf B399} (1993) 211\\
                J.\ Gegenberg and G.\ Kunstatter, Phys.\ Rev.\ {\bf D47} (1993) R4192\\
                K.\ Kuchar, Geometrodynamics of Schwarzschild black holes, prep.\ 
                UU--Rel--94--3--1
\bibitem[2]{tei}B.M.\ Barbashov, V.V. Nesterenko, A.M.\ Chervjakov, 
                Theor.\ Mat.\ Phys.\ {\bf 40} (1979) 15; 
                E.\  Teitelboim, Phys.\ Lett {\bf B126} (1983) 41; E.\ d'Hooker, D.Z.\ 
                Freedman and R.\ Jackiw, Phys.\ Rev.\ {\bf D28} (1983) 1583; R.\ Jackiw, 
                Jucl.\ Phys.\ {\bf B252} (1985) 343; C.\ Cangemi and R.\ Jackiw, Phys.\ 
                Rev.\ Lett.\ {\bf 69} (1992) 233
\bibitem[3]{bro}J.D.\ Brown, M.\ Henneaux and C.\ Teitelboim, Phys.\ Rev.\ {\bf D31} 
                (1986) 319; A.\ Mandal, A.M.\ Sengupta and S.R.\ Nada, Mod.\ Phys.\ Lett.\ 
                {\bf A6} (1991) 1685; E.\ Witten, Phys.\ Rev.\ {\bf D44} (1991) 314; 
                C.G.\ Callan, S.P.\ Giddings, J.A.\ Harvey and A.\ Strominger, Phys.\ 
                Rev.\ {\bf D45} (1992) R1005
\bibitem[4]{ban}D.\ Banks and M.\ O'Loughlin, Nucl.\ Phys.\ {\bf B362} (1991) 649; S.D.\ 
                Odintsov and I.J.\ Shapiro, Phys.\ Lett.\ {\bf B263} (1991) 183 and 
                Mod.\ Phys.\ Lett.\ {\bf A7} (1992) 437; I.G.\ Russo and A.A.\ Tseytlin, 
                Nucl.\ Phys.\ {\bf B382} (1992) 259;\ Volovich, Mod.\ Phys.\ Lett.\ A (1992) 1827; R.P.\ 
                Mann, Phys.\ Rev.\ {\bf D47} (1993) 4438; D.\ Louis--Martinez, J.\ 
                Gegenberg and G.\ Kunstatter, Phys.\ Lett.\ {\bf B321} (1994) 193; D.\ 
                Louis--Martinez and G.\ Kunstatter, Phys.\ Rev.\ {\bf D49} (1994) 5227
\bibitem[5]{man}R.B.\ Mann, A.\ Shiekh and L.\ Tarasov, Nucl.\ Phys.\ {\bf B341} (1990) 
                134
\bibitem[6]{lem}J.S.\ Lemos and  P.M.\ Sa, Phys.\ Rev.\ {\bf D49} (1994) 2897
\bibitem[7]{kata}M.O.\ Katanaev and I.V.\ Volovich, Phys.\ Lett.\ {\bf B175} (1986) 413, 
                Am.\ Phys.\ (N.Y.) {\bf 197} (1990) 1; 
                M.O.\ Katanaev, J.\ Math.\ Phys.\ {\bf 31} (1990) 882 and {\bf 32} 
                (1991) 2483
\bibitem[8]{katb}M.O.\ Katanaev, J.\ Math.\ Phys.\ {\bf 34} (1993) 
                700
\bibitem[9]{kuma}W.\ Kummer and D.J.\ Schwarz, Phys.\ Rev.\ {\bf D45} (1992) 3628
\bibitem[10]{kumb}W.\ Kummer and D.J.\ Schwarz, Nucl.\ Phys.\ {\bf B382} (1992) 171; F.\ 
                Haider and W.\ Kummer, Journ.\ of Mod.\ Phys.\ {\bf A9} (1994) 207
\bibitem[11]{sch}P.\ Schaller and T.\ Strobl, Class.\ Quant.\ Grav.\ {\bf 11} (1993) 331
\bibitem[12]{stra}T.\ Strobl, Phys.\ Rev.\ {\bf D50} (1994) 7356;
                P.\ Schaller and T.\ Strobl, Mod.\ Phys.\ Lett.\ {\bf A9} 
                (1994) 3129; P.\ Schaller and T.\ 
                Strobl, Quantization of field theories generalizing gravity -- 
                Yang--Mills systems on the cylinder, Prep.\ TUW--94--02, Proceedings 
                of the Baltic RIM Student Seminar (to be published in 'Lecture Notes 
                in Physics')
\bibitem[13]{strb}T.\ Strobl, Poisson--structure induced field theory and models of 1+1 
                dimensional gravity, PhD--thesis, TU Wien, June 1994
\bibitem[14]{int}In the quantum case the $C_i$ are identified for nontrivial topology on 
                the space $R \times S^1$  as the only genuine variables of the 
                theory [11,12]
\bibitem[15]{klo}T.\ Kl\"osch and T.\ Strobl, Cl.\ Quantum Grav.\ 
                {\bf 13} (1996) 965 and 1191  
\bibitem[16]{kumc}W.\ Kummer and P.\ Widerin, Mod.\ Phys.\ Lett {\bf A9} (1994) 1407
\bibitem[17]{yor}J.W.\ York, Found.\ Phys.\ {\bf 16} (1986) 249; J.D.\ Brown and J.W.\ 
                York, Phys.\ Rev.\ {\bf D47} (1993) 1407
\bibitem[18]{wal} R.M.\ Wald, Phys.\ Rev.\ 
                {\bf D48} (1993) R3427; V.\ Iyer and R.M.\ Wald, Phys.\ Rev.\ {\bf D50} 
                (1994) 846, for a recent comparison of that approach of one 
                of \cite{yor} cf. V.\ Iyer and R.M.\ Wald, a comparison of 
                Noether charge and Euclidean methods for computing the 
                entropy of stationary black holes, Univ.\ of Chicago prep., 
                gr--qc--9503052
\bibitem[19]{kume} W.\ Kummer and P.\ Widerin, Phys.\ Rev.\ D 
                {\bf 52} (1995) 6965
\bibitem[20]{katc}M.O.\ Katanaev, W.\ Kummer, H.\ Liebl, 
                Phys.\ Rev.\ D {\bf 53} (1996) 5609
\bibitem[21]{he}  R.\ Hecht, F.W.\ Hehl, J.D.\ McCrea, E.W.\ Mielke, Y.\ Ne'eman, 
                 Phys. Lett. {\bf A172} (1992) 13; F.W.\ Hehl, J.D.\ 
                 McCrea, E.W.\ Mielke, Y.\ Ne'eman, {\sl 
                Metric--affine gauge theory of gravity field 
                equations, Noether identities, world spinors, and 
                breaking of dilaton invariance}, Physics Reports 
                1995 (to be published)
\bibitem[22]{kumd}W.\ Kummer, Deformed Iso--(2,1)--symmetry and non--Einsteinian 
                2d--gravity with matter, Hadron Structure '92 (Eds.\ D.\ Brunsko and J.\ 
                Urb\'an) p.\ 48--56, Ko\v sice 1992; \\
                W.\ Kummer, Exact classical and quantum integrability of $R^2+T^2$ 
                gravity in 1+1 dimensions, Proc.\ Int.\ Europhys.\ Conf.\ on High 
                Energy Physics (Eds.\ J.\ Carr and M.\ Perrottet) p.\ 251--253, Edition 
                Fronti\'eres, Gif-sur-Yvette 1994

\end{thebibliography}
\end{document}